\documentclass[lettersize,journal]{IEEEtran}
\usepackage{amsmath,amsfonts}
\usepackage{algorithmic}
\usepackage{algorithm}
\usepackage{array}
\usepackage[caption=false,font=normalsize,labelfont=sf,textfont=sf]{subfig}
\usepackage{textcomp}
\usepackage{stfloats}
\usepackage{url}
\usepackage{verbatim}
\usepackage{graphicx}
\usepackage{booktabs}
\usepackage{multirow}
\usepackage{graphicx} 
\usepackage{hyperref}
\usepackage{cite}

\begin{document}

\title{Rethinking the Upsampling Layer in Hyperspectral Image Super Resolution}

\author{\IEEEauthorblockN{Haohan Shi, Fei Zhou,
Xin Sun, ~\IEEEmembership{Member,~IEEE,}
Jungong Han, ~\IEEEmembership{Senior Member,~IEEE}
}
\thanks{H. Shi and X. Sun are with Faculty of Data Science, City University of Macau, 999078, SAR Macao, China. F. Zhou is with College of Oceanography and Space Informatics, China University of Petroleum (East China), Qingdao, China. J. Han is with Department of Automation, Tsinghua University, Beijing, China. } 
\thanks{This work is supported by the Science and Technology Development Fund, Macao SAR No.0006/2024/RIA1 and National Natural Science Foundation of China under Project No.61971388.}}

\markboth{Journal of \LaTeX\ Class Files,~Vol.~14, No.~8, August~2025}%
{Shell \MakeLowercase{\textit{et al.}}: A Sample Article Using IEEEtran.cls for IEEE Journals}


\maketitle

\begin{abstract}

Deep learning has achieved significant success in single hyperspectral image super-resolution (SHSR); however, the high spectral dimensionality leads to a heavy computational burden, thus making it difficult to deploy in real-time scenarios. To address this issue, this paper proposes a novel lightweight SHSR network, i.e., LKCA-Net, that incorporates channel attention to calibrate multi-scale channel features of hyperspectral images. Furthermore, we demonstrate, for the first time, that the low-rank property of the learnable upsampling layer is a key bottleneck in lightweight SHSR methods. To address this, we employ the low-rank approximation strategy to optimize the parameter redundancy of the learnable upsampling layer. Additionally, we introduce a knowledge distillation-based feature alignment technique to ensure the low-rank approximated network retains the same feature representation capacity as the original. We conducted extensive experiments on the Chikusei, Houston 2018, and Pavia Center datasets compared to some SOTAs. The results demonstrate that our method is competitive in performance while achieving speedups of several dozen to even hundreds of times compared to other well-performing SHSR methods.

\end{abstract}


\section{Introduction}
\label{intro}
Hyperspectral imaging captures hundreds or even thousands of contiguous spectral bands with very high resolution. It provides discriminative spectral information to distinguish substances with quite similar appearances  \cite{hong2024spectralgpt,du2023exploring,gao2024appearance}. Therefore, hyperspectral imaging has been widely applied in various tasks, including target detection \cite{targetdetection, targetdetection2, targetdetection3, targetdetection4} and tracking \cite{targettracking, sun2020deep}, change detection \cite{changedetection}, semantic segmentation \cite{hong2023cross, 10750822} and land cover classification \cite{classification2, classification3, zhou2023adaptive}. However, there is an inherent trade-off between spatial and spectral resolution due to physical constraints \cite{toward}. That is, hyperspectral imaging devices typically enlarge the size of photodiodes or photosensitive elements to effectively capture the spectral information of each pixel at different wavelengths. It reduces the total number of pixels in the sensor, resulting in low spatial resolution of the image. This obstacle hinders hyperspectral images in some critical high-precision remote sensing applications.
\begin{figure}[t]
  \centering
  \begin{minipage}{0.248\textwidth}
    \centering
    \hspace{-2mm}\includegraphics[height=3.4cm]{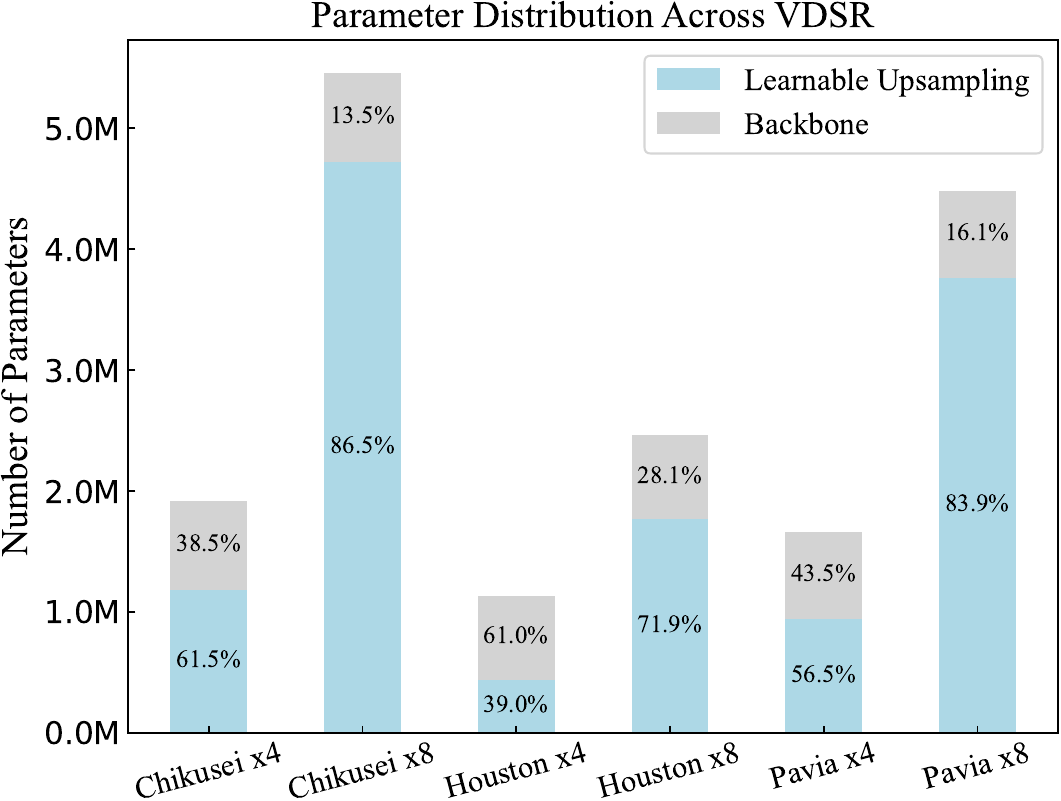}
  \end{minipage}%
  \begin{minipage}{0.248\textwidth}
    \centering
    \hspace{-2mm}\includegraphics[height=3.4cm]{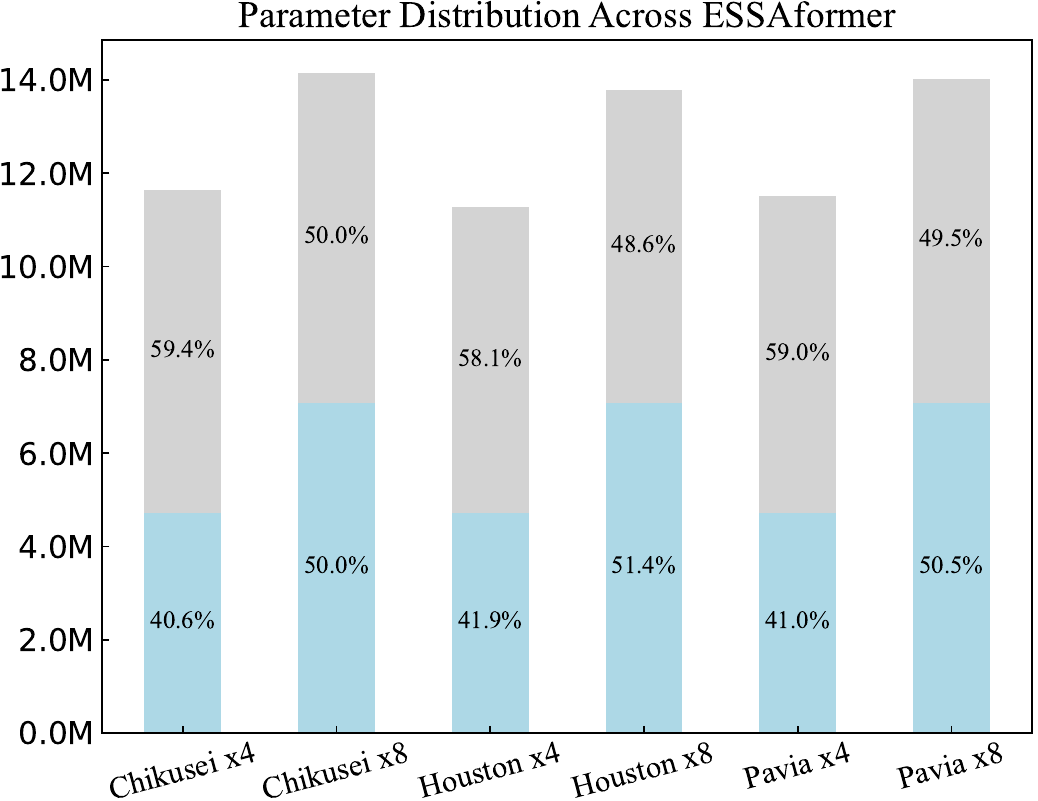}
  \end{minipage}
  \caption{Illustration of the number of parameters for different functional layers of VDSR  \cite{vdsr} and ESSAfomer \cite{essaformer}. Blue represents the proportion of the learnable upsampling layer's parameters, while gray represents the backbone.}
  \label{occupation}
\end{figure}

To address the above challenge, recent work has paid attention to the hyperspectral image super-resolution (HISR) area. For example, multispectral and hyperspectral image fusion (MHF) \cite{mhf1, mhf2} is studied to generate high spatial resolution hyperspectral images complemented with multispectral cameras. Hyperspectral images provide more detailed information than multispectral images, which is able to accurately distinguish closely related materials. However, MHF relies heavily on auxiliary images, which demands low-resolution multispectral and high-resolution hyperspectral images to be captured in the same scene for precise registration. It is difficult to meet such requirements in practical scenarios. Therefore, single hyperspectral image super-resolution (SHSR) methods have gained significant traction in recent years. Deep learning methods, such as convolutional neural networks (CNNs) and Transformers \cite{attention} are typically employed to learn the mapping between low-resolution and high-resolution images for generating high-quality super-resolution images. These methods effectively bypass the challenges associated with image registration and the complexities of data fusion.

SHSR methods have significantly reduced computational costs compared to MHF methods. However, they still face challenges with high complexity when extracting spectral features using deep learning techniques \cite{gpu}. Lightweight SHSR models are in high demand for real-time applications, such as drones and satellites. Most existing SHSR methods consist of a feature extraction backbone and a learnable upsampling layer. Recent lightweight approaches primarily focus on optimizing the backbone through techniques like channel separation, lightweight convolutions, attention modules, and structural re-parameterization. However, our findings reveal that, particularly in hyperspectral super-resolution tasks, the learnable upsampling layer contributes the most to computational complexity. For instance, as illustrated in Fig. \ref{occupation}, we compare the parameter distribution between the backbone and the learnable upsampling layers of two popular super-resolution networks: VDSR \cite{vdsr} and ESSAformer \cite{essaformer}.

This paper investigates lightweight SHSR networks, and introduces a novel network structure, LKCA-Net, based on Large-Kernel Channel Attention and the learnable upsampling layer. To address the lightweight challenges of the learnable upsampling layer, we explore effective optimization strategies. Specifically, we utilize Singular Value Decomposition (SVD) to reveal the low-rank property of the learnable upsampling layer and examine how low-rank approximation technique affect super-resolution performance. Additionally, we emphasize the importance of a feature alignment strategy in enhancing the performance of models with low-rank approximation applied to the upsampling layer.

Notably, our method can optimize the conventional structure of the learnable upsampling layers in most of the deep networks. This indicates the proposed optimization method is non-intrusive and can be easily integrated into existing networks without changing the main structure of the model itself. Overall, the major contributions of this work can be summarized as follows:
\begin{enumerate}
  \item This paper presents a novel SHSR model, i.e., LKCA-Net, aiming to calibrate the multi-scale features of spectral channels while minimizing the computational cost for hyperspectral image super-resolution.
  \item This paper proposes, for the first time, that the low-rank property of the learnable upsampling layer is critical for model lightweight in the SHSR task, and confirms this hypothesis with low-rank approximation.
  \item This paper further introduces a feature alignment strategy based on knowledge distillation to preserve the feature representation capability of the learnable upsampling layer in a low-rank approximated network, ensuring it remains consistent with that of the original network.
  \item Extensive experimental results demonstrate the efficacy of our LKCA-Net on the Chikusei, Houston2018, and Pavia datasets compared to SOTAs. The results highlight the significance of our contributions to low-rank approximation and the feature alignment strategy for the learnable upsampling layer. 
\end{enumerate}

The rest of this article is organized as follows: Section \ref{RW} briefly introduces the related work. Section \ref{method} proposes one network LKCA-Net, discusses the low-rank nature of the upsampling layer, and suggests a feature alignment strategy. Section \ref{exp} shows the experimental results. Section \ref{con} finally concludes the article.

\section{Related Work}
\label{RW}
This section first briefly describes recent developments in related fields, including single hyperspectral image super-resolution and related lightweight network models.
\subsection{Single Hyperspectral Image Super-resolution}
SHSR techniques typically employ deep learning models to learn the mapping between low-resolution and high-resolution images, producing high-quality images. Such an approach eliminates the need for auxiliary images, thereby avoiding complex image registration tasks. Deep learning models, such as CNNs and attention mechanisms, are now widely utilized as SHSR feature extractors.

Early SHSR models were mainly constructed based on CNNs. For example, the 3D-FCNN model \cite{3dfcnn} introduced 3D convolution into SHSR, enabling simultaneous extraction of spatial and spectral features while preserving spectral correlations. SRDNet \cite{srdnet} proposed a hybrid convolutional structure that combined 2D and 3D convolution units, leveraging 2D convolutions to focus on spatial feature extraction and 3D convolutions to capture spectral information. In recent years, attention mechanisms have become increasingly popular in constructing backbone structures for SHSR models. And channel attention \cite{ca} is utilized to adaptively adjust the feature weights of each channel to better capture interdependence. The SSPSR network \cite{sspsr} incorporated both spatial and spectral attention residual modules to capture the correlations between spatial and spectral context within hyperspectral images. SGARDN \cite{sgardn}, leveraged group convolutions and spectral attention mechanisms to capture shallow spatial-spectral features. Some works integrated both spatial and spectral attention. For instance, MSSR network \cite{mssr} utilized a dilated convolution-based attention to expand the receptive field within a shallow network, so as to capture global spatial information. Meanwhile, SRDNet \cite{srdnet} introduced a pyramid-structured self-attention mechanism to model spectral features in both spatial and spectral contexts. KNLConv \cite{knlconv} overcomes the limitations of standard convolution by exploring non-local dependencies in the kernel space.

The work of “Attention Is All You Need” \cite{attention} proliferates the employment of Transformers into SHSR models. MSDformer \cite{msdformer}, for instance, combined the strengths of CNNs for local information extraction with the Transformer’s capability for capturing global information. Although the Transformer structure has many advantages in feature extraction, the high computational cost for hyperspectral images is a critical concern. Therefore, Interactformer \cite{interactformer} replaced the self-attention with separable self-attention in the Transformer to reduce memory consumption. ESSAformer \cite{essaformer} introduced kernelized self-attention mechanism to transform the spectral correlation coefficient into radial basis function kernel, which approximated the similarity between Query (\textit{Q}) and Key (\textit{K}) with an exponential mapping.

\subsection{Lightweight Networks for SHSR}
Due to the high spectral dimensionality of a hyperspectral image, which contains a massive amount of information, deep learning involves a large number of parameters and substantial computational cost to extract features \cite{10817590, 3681270}. Additionally, the upsampling commonly used in the super-resolution process further increases the computational complexity. Therefore, it arises high demands for lightweight networks to achieve SHSR tasks on resource-constrained environments.

In the field of remote sensing, a few approaches focused on lightweight super-resolution models by designing lightweight feature extractors \cite{sspsr, msdformer, fenet}. The common way is to divide the input image along the channel dimension into multiple groups, which are then processed by different branches within the model. For instance, FetNet \cite{fenet} split the input features into two branches, each of which only processed half of the features to reduce computational complexity. MRFN \cite{mrfn} is a lightweight multi-resolution feature fusion network that employed PixelUnshuffle \cite{pixel} to downsample the input image to multiple resolutions, creating inputs for different scale branches. Additionally, MRFN used cheap convolutions \cite{cheapconv} and lightweight channel attention (LCA) \cite{lca} to further reduce computational complexity.
\begin{figure*}[t]
	\centering
	\includegraphics[scale=0.58]{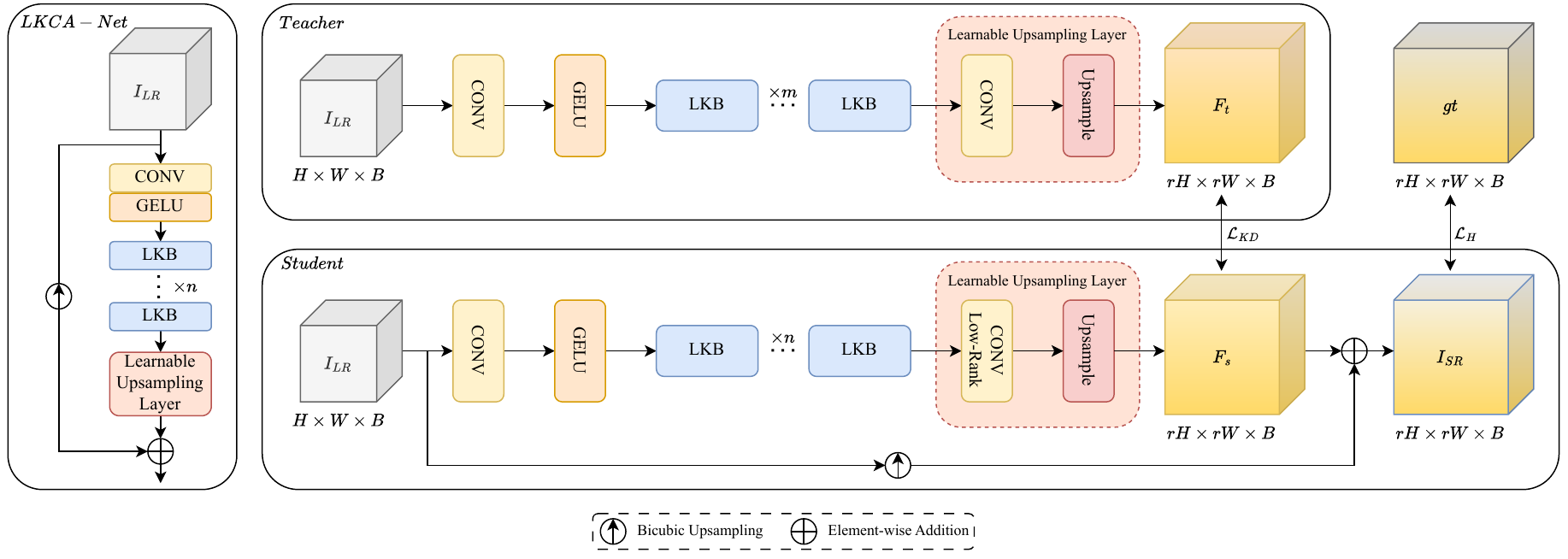}
    \caption{Overall Architecture of  LKCA-Net}
	\label{net}
\end{figure*}

The efficacy of group convolution to reduce computational complexity was demonstrated in SSPSR \cite{sspsr} and SGARDN \cite{sgardn} for optimizing convolutional and attention modules. Interactformer \cite{interactformer} introduced a separable self-attention module with linear complexity to lower memory consumption, as opposed to the quadratic complexity of traditional self-attention. ESSAformer \cite{essaformer} incorporated kernelized self-attention and made improvements in computational efficiency. Lightweight convolutional kernels were also commonly used. RFNet \cite{RFSR} and SRDNet \cite{srdnet} decomposed 3D convolutional kernels (3x3x3) into two separate kernels, focusing on spatial (3x3x1) and spectral (1x1x3) dimensions, which reduced the computational burden of full 3D convolution. MSSR \cite{mssr} further reduced the computational cost by retaining 3D convolutions only along the spectral dimension. Guo et al. \cite{toward} proposed a lightweight approach for hyperspectral super-resolution with explicit degradation estimation. It mimicked spatial and spectral degradation by anisotropic Gaussian kernels \cite{Gaussiankernels} and a mixed Gaussian model \cite{mixedGaussian, sun2021gaussian} to accurately capture the degradation process with fewer parameters. This significantly reduced the parameters and computational complexity, making it highly suitable for resource-constrained environments. Some works employed structural reparameterization for lightweight. For example, RepCPSI\cite{repcpsi} combined two reparameterization methods \cite{acnet, repvgg} with a multi-branch structure to extract diverse feature categories, then merged them into a single-path structure for inference.

Currently, lightweight methods focused on the backbone have been studied \cite{10496826}, yet the issue of parameter redundancy still exists within models. And the upsampling layer seems to have been overlooked. This paper not only focuses on a new lightweight network, but also solve the problem from the viewpoint of optimizing the learnable upsampling layer.

\section{PROPOSED METHOD}
\label{method}

\begin{figure*}[ht]
	\centering
	\includegraphics[scale=0.58]{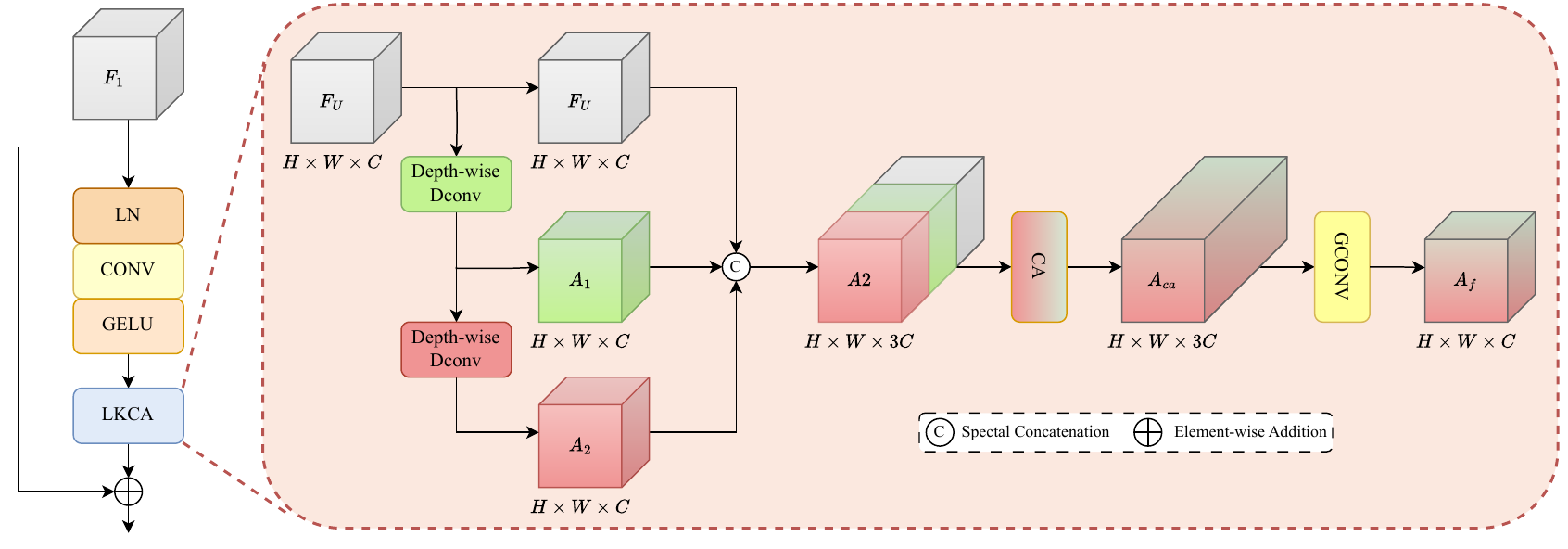}
    \caption{Large-Kernel Channel Attention-based Block}
	\label{lkb}
\end{figure*}

\subsection{Overall Architecture and Problem Definition}
In this paper, we first design a lightweight SHSR model with Large-Kernel Channel Attention, named LKCA-Net. LKCA-Net features a concise architecture, primarily a backbone composed of stacked LKCA-based Block (LKB) modules to extract deep features, a learnable upsampling layer to reconstruct spatial resolution, and a bicubic upsampling connection to boost performance. Fig. \ref{net} illustrates the overall architecture of the network. Moreover, we argue that the computational complexity of SHSR models can be significantly reduced by optimizing the upsampling process. To validate this hypothesis, we further conduct the low-rank approximation on the upsampling layer, which we will discuss and confirm later.

Specifically, the low-resolution hyperspectral image is indicated as $I_{LR} \in \mathbb{R}^{H \times W \times B}$, the high-resolution hyperspectral image is denoted as $I_{HR} \in \mathbb{R}^{rH \times rW \times B}$, and the reconstructed result is $I_{SR} \in \mathbb{R}^{rH \times rW \times B}$. Here, $H$ and $W$ represent the spatial resolution of the image, $B$ denotes the number of spectral bands, and $r$ is the scaling factor. The overall architecture can be expressed by the following equation:
\begin{equation}
\label{se}
I_{SR} = H_{UP}(H_{SR}(I_{LR})) + F_{Bicubic},
\end{equation}
where $H_{UP}$ denotes learnable upsampling layer and the $F_{Bicubic} \in \mathbb{R}^{rH \times rW \times B}$ denotes the bicubic interpolation residual. $H_{SR}(\cdot)$ is our LKCA-Net backbone, to clarify, $H_{SR}$ consists with one $3 \times 3$ convolution to extract shallow features $F_0$, and $N$ stacked LKB modules to extract hierarchical features. This procedure can be expressed as:
\begin{equation}
F_0 = Conv_{3 \times 3}(I_{LR}), 
\end{equation}
\begin{equation}
F_N = H_{LKB,N}(H_{LKB,N-1}(\cdots  (H_{LKB,1}(F_0))\cdots)),
\end{equation}
where $F_N \in \mathbb{R}^{H \times W \times C}$ represents the feature map extracted by the backbone. $H_{LKB,N}(\cdot)$ denotes the $N^{th}$ LKB module in the backbone. Subsequently, $F_N$ passes through a learnable upsampling layer $H_{UP}$ which consists of a $3 \times 3$ convolution and a Pixelshuffle upsampling to reconstruct high-resolution $F_{UP} \in \mathbb{R}^{rH \times rW \times B}$. This process can be expressed as
\begin{equation}
\label{se}
F_{UP} = H_{UP}(F_N).
\end{equation}

Finally, $F_{UP}$ is fused with the high-resolution feature map obtained through Bicubic interpolation, $F_{Bicubic} \in \mathbb{R}^{rH \times rW \times B}$, to reconstruct the final high-resolution hyperspectral image $I_{SR}$. This process can be expressed as
\begin{equation}
\label{se}
F_{Bicubic} = H_{B}(I_{LR}),
\end{equation}
\begin{equation}
\label{se}
I_{SR} = F_{UP} + F_{Bicubic},
\end{equation}
where, $H_B(\cdot)$ represents the definition function of Bicubic upsampling. The final output of the network is a hyperspectral image with $B$ spectral bands and a spatial resolution of $rH \times rW$. 

In order to maintain the same feature representation capability for the low-rank approximated network with the original, we employed a feature alignment strategy based on knowledge distillation to assist in training the low-rank approximated network. The process of feature alignment is illustrated in Fig. \ref{net}. We use a low-rank approximated network with $n$ LKB modules as the student network, and a network with $m$ LKB modules ($m > n$) of high representation capability as the teacher network. The feature maps after PixelShuffle are used as knowledge to be transferred between the networks. We utilize knowledge distillation loss to minimize the difference between the student feature maps $F_s$ and $F_t$ of the teacher. It allows the student to mimic the upsampling results of the teacher during training, thus achieving performance comparable to the teacher network.

\subsection{Large-Kernel Channel Attention-based Block}
The LKB module serves as the basic block of the LKCA-Net backbone, which is designed to extract spatial-spectral features. The structure of the LKB is illustrated in Fig. \ref{lkb}. The LKB consists of an $LN-Conv_{1 \times 1}-GELU$ block for feature projection, one LKCA module to capture multi-scale features, and a residual connection to facilitate information flow. We design the LKCA module with multi-scale features and channel-wise attention to form an efficient and lightweight network for SHSR tasks to extract complex features. Specifically, the LKB module can be expressed as:
\begin{equation}
\label{se}
F_{LKB} = H_{LKB}(F_1),
\end{equation}
where $F_1$ and $F_{LKB} \in \mathbb{R}^{H \times W \times C}$ represent the input and output feature maps, and $H_{LKB}(\cdot)$ denotes the LKB module.

The LKCA module adopts convolution decomposition strategies to expand the receptive field while reducing computational complexity. We construct a cascade of two depthwise dilated convolutions with different dilation rates, denoted as ($Conv_{k1}^{d1}-Conv_{k2}^{d2}$), where $k1$ and $k2$ represent the kernel sizes, and $d1$ and $d2$ are dilation rates. This strategy not only enlarges the receptive field but also naturally generates multi-scale features. The outputs of these two dilated convolutions are concatenated to fuse the multi-scale features.

One critical problem is that the depthwise convolution processes of each channel are independent, which may limit the exchange of information between channels. However, large redundant features exist within multi-scale channels. To overcome this limitation, we incorporate group convolution and a channel attention mechanism to calibrate the multi-scale channel features. The CA mechanism effectively captures dependencies between channels and improves interactions among information. It dynamically assigns weights to the channels of the feature map, adjusting the importance of each channel, helping the model to learn and restore relationships between channels.  In the final integration stage, we further replace the original $1 \times 1$ convolution with grouped convolution to further reduce the parameters of the module. The process can be expressed as:
\begin{equation}
\label{se}
F_W = H_{LKCA}(F_U),
\end{equation}
where $F_U$ and $F_W \in \mathbb{R}^{H \times W \times C}$ represent the input and output feature maps respectively. $H_{LKCA}(\cdot)$ denotes the LKCA module. In LKCA, $F_U$ is first passed through a dilated convolution with a dilation rate of $5$, yielding the first attention map $A_1 \in \mathbb{R}^{H \times W \times C}$. Then, $A_1$ is passed through another dilated convolution with a dilation rate of $7$, producing the second attention map $A_2 \in \mathbb{R}^{H \times W \times C}$. Subsequently, $F_U$, $A_1$, and $A_2$ are concatenated along the channel dimension to form a combined attention map $A_C \in \mathbb{R}^{H \times W \times 3C}$. This process can be expressed as:
\begin{equation}
\label{se}
A_1 = Conv^{5}(F_U), 
\end{equation}
\begin{equation}
\label{se}
A_2 = Conv^{7}(A_1), 
\end{equation}
\begin{equation}
\label{se}
A_C = \text{Cat}(F_U, A_1, A_2),
\end{equation}
where $Conv^{5}(\cdot)$ and $Conv^{7}(\cdot)$ represent the dilated convolutions with dilation rates of $5$ and $7$, respectively. Afterwards, $A_C$ is passed through the CA module to obtain the attention map $A_{ca} \in \mathbb{R}^{H \times W \times 3C}$, where channel-wise redundancy is reduced. Finally, $A_{ca}$ is processed by a $1 \times 1$ group convolution and integrated into the final attention map $A_f \in \mathbb{R}^{H \times W \times C}$. This process is expressed as:
\begin{equation}
\label{se}
A_{ca} = H_{CA}(A_C), 
\end{equation}
\begin{equation}
\label{se}
A_f = H_g(A_{ca}),
\end{equation}
where $H_{CA}(\cdot)$ represents the function definition of the CA module, and $H_g(\cdot)$ denotes the $1 \times 1$ group convolution. Finally, $A_f$ is multiplied element-wise with $F_U$ to obtain the feature map $F_W$ with enhanced attention, as shown below:
\begin{equation}
\label{se}
F_W = A_f \ast F_U.
\end{equation}

The final output of the LKB module contains rich spatial and spectral features, enabling our model to achieve promising SHSR performance.

\subsection{Upsampling Layer is Low Rank}

In this section, we demonstrate that the parameter matrix of the learnable upsampling layer in SHSR models is low-rank. During the upsampling stage of the SHSR model, the PixelShuffle method requires the channels to satisfy Eq. \ref{pixelshuffle}.
\begin{equation}
\label{pixelshuffle}
C_{out} = C \times r^2,
\end{equation}
where, $C$ is the number of spectral channels in the hyperspectral image, and $r$ is the scale factor. Therefore, the number of parameters for the convolution layer before the PixelShuffle is $C_{in} \times C \times r^2 \times k^2$, where $k$ is the kernel size of the convolution. 
\begin{figure}[tbp]
	\centering
	\includegraphics[scale=0.40]{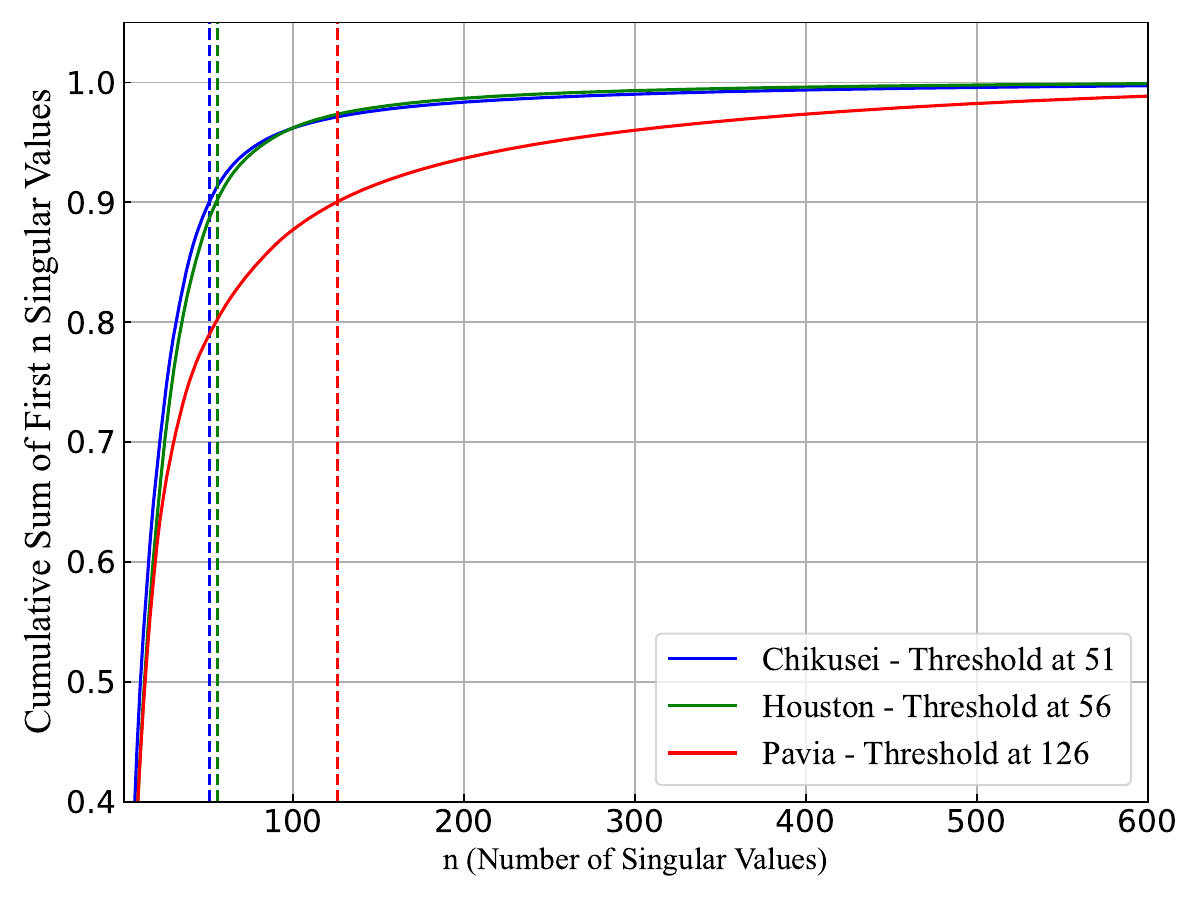}
	\caption{Cumulative Sum of Normalized Singular Values. }
	\label{SVD}
\end{figure}

There is no doubt that the parameter matrix of the learnable upsampling layer is a high-dimensional tensor. The parameters of the upsampling layer grow exponentially with the scale factor and kernel size. Therefore, we have to investigate the inherent property of this high-dimensional tensor of the upsampling layer. At first, we reshape the parameter tensor into a two-dimensional matrix $M \in \mathbb{R}^{C_{out}, C_{in} \times k^2}$ and then perform singular value decomposition (SVD) on the matrix to analyze the rank of the parameter matrix.

For our model, the value of $C_{in}$ is $128$, and the kernel size is $3$. After reshaping, the parameter tensor of the learnable upsampling layer should become a matrix of size $[C_{out}, 3 \times 3 \times 128]$, meaning that in the case of a full rank, the rank of matrix $M$ should be 1152. Subsequently, we perform SVD decomposition on the matrix $M$, as shown in Equ. \ref{svd}.

\begin{small}
\begin{equation}
\label{svd}
M = U \Sigma V^T =
\begin{bmatrix}
u_1&u_2&\cdots &u_m
\end{bmatrix}
\begin{bmatrix}
\sigma_1 & 0 & \cdots & 0 \\
0 & \sigma_2 & \cdots & 0 \\
\vdots & \vdots & \ddots & \vdots \\
0 & 0 & \cdots & \sigma_p \\
\vdots & \vdots &  & \vdots \\
0 & 0 & \cdots & 0 \\
\end{bmatrix}
\begin{bmatrix}
v_1^T \\
v_2^T \\
\vdots \\
v_n^T
\end{bmatrix},
\end{equation}
\end{small}

\noindent where $\sigma_1, \sigma_2, \cdots, \sigma_p$ are all singular values of the matrix $M$. After normalizing these singular values, Fig. \ref{SVD} illustrates the cumulative distribution of singular values of the matrix on three datasets with one $\times 4$ scale factor. The y-axis represents the cumulative sum of the top-$n$ singular values. The x-axis represents the indices of the singular values, with the largest singular value positioned on the far left and decreasing sequentially. The total number of indices should be 1152, and we only show the cumulative distribution of the first 600 singular values.

From the curves of Fig. \ref{SVD}, it can be observed that the cumulative sum of singular values rapidly approaches $1$ as the singular value index increases. The distribution of singular values exhibits a clear long-tail characteristic. This indicates that the majority of the information is concentrated in a small number of larger singular values, which demonstrates the low-rank property of the matrix. Therefore, we can approximate the matrix $M$ via low-rank approximation.

\begin{figure}[htbp]
	\includegraphics[scale=0.465]{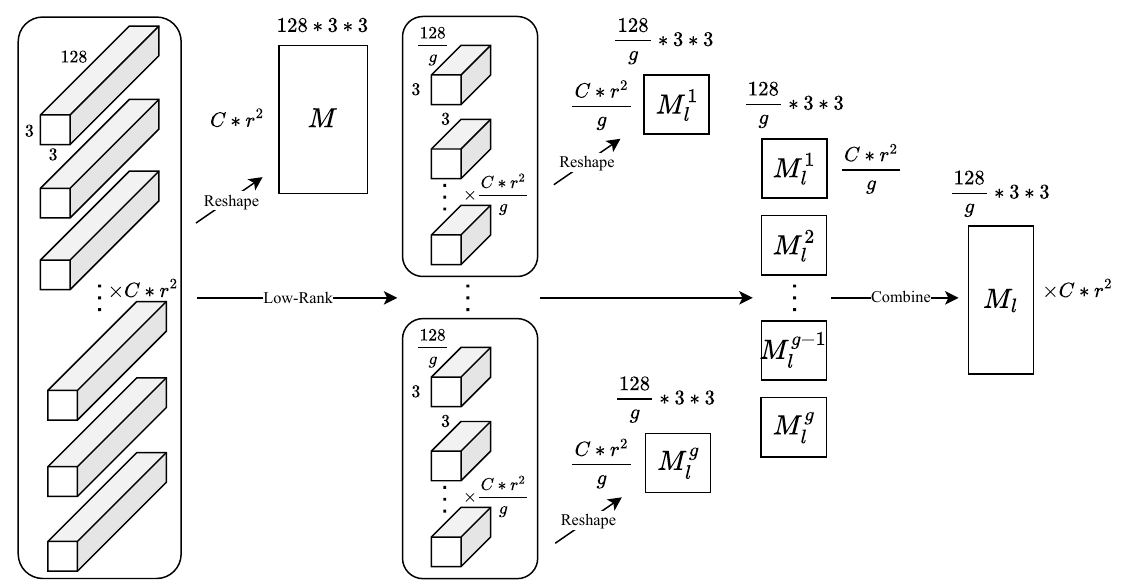}
    \caption{The original $3\times3$ convolution is approximated by a group convolution with $g$ groups, reducing the number of parameters by $g$. The low-rank matrix $M_l$ is formed by reshaping the parameters of the group convolution.}
	\label{low-rank}
\end{figure}

We introduce the group convolution to approximate the matrix $M$ as shown in Fig. \ref{low-rank}. It illustrates the process of approximating the matrix $M$ with a low-rank matrix. Clearly, the factor $g$ significantly affects the trade-off between network performance and complexity. As $g$ must be selected from integers divisible by $r^2$ to satisfy the PixelShuffle demand, $g \in (2, 4, 8, 16)$ when $r=4$. We can observe from Fig. \ref{SVD} that the cumulative sum of singular values reaches 90\% with about the first 60 singular values on the Chikusei and Houston datasets. Therefore, a rank-60 matrix is sufficient to effectively approximate the parameter matrices. For the Pavia dataset, the parameter matrix requires a rank-130 matrix to achieve a good approximation. In general, we ultimately set $g$ to $8$.

\subsection{Feature Alignment}
\label{FA}

To address the performance gap caused by the low-rank approximation for the upsampling layer, we propose a feature alignment strategy based on knowledge distillation (KD). We design one feature alignment loss $\mathcal{L}_{total}$ to align the features between the approximated layer of the student network and the teacher network. $\mathcal{L}_{total}$ consists of two components: the supervised loss $\mathcal{L}_{H}$ and knowledge distillation loss $\mathcal{L}_{KD}$. 

At first, we adopt the $H$ loss suggested by MSDformer \cite{msdformer} as the supervisory loss $\mathcal{L}_{H}$. $H$ loss comprehensively considers the reconstruction loss $\mathcal{L}_1$, spectral angular loss $\mathcal{L}_{sam}$, and gradient loss $\mathcal{L}_{grad}$. The definition of $H$ loss could be formulated as
\begin{equation}
\label{se}
\mathcal{L}_{H}(\Theta) =\mathcal{L}_1 + \lambda_1 \mathcal{L}_{sam} + \lambda_2 \mathcal{L}_{grad},
\end{equation}
where, $\lambda_1$ and $\lambda_2$ are the weight coefficients for spectral loss and gradient loss, respectively, used to balance the contributions of different loss components. We set $\lambda_1 = 0.5$ and $\lambda_2 = 0.1$ as suggested in \cite{msdformer}.

For the $KD$ loss, we emphasize spectral consistency for SHSR tasks by adopting the $SAM$ loss, allowing the student network to learn the spectral features of the teacher network. We also include $grad$ loss for spatial features. Given the significance of spectral consistency in the SHSR task, cosine similarity loss ($cos$ loss) is employed to further preserve this consistency. The loss function is defined as
\begin{equation}
\mathcal{L}_{cos}(\Theta) = 1 - \frac{1}{N} \sum_{n=1}^N \cos \left( \frac{H_{tr}^n \cdot H_{sr}^n}{\|H_{tr}^n\|_2 \cdot \|H_{sr}^n\|_2} \right),
\end{equation}
where, $N$ represents the number of images in a training batch, $H_{tr}^n$ is the high-resolution image reconstructed by the teacher net for the $n^{th}$ image in the training batch, and $H_{sr}^n$ is the high-resolution image reconstructed by the student. $\|\cdot\|_2$ represents the $\ell_2$ norm. Like $SAM$ loss, $cos$ loss focuses on optimizing spectral directions without considering the magnitude, but it calculates cosine similarity for each pixel's spectral vectors, aligning student and teacher spectral directions. While $SAM$ loss has stronger physical significance, the $cos$ loss is simpler, more efficient, and easier to optimize. Finally, the $KD$ loss is defined as:

\begin{equation}
\mathcal{L}_{KD}(\Theta) = \lambda_3 \mathcal{L}_{cos} + \lambda_4 \mathcal{L}_{sam} + \lambda_5 \mathcal{L}_{grad},
\end{equation}
where $\lambda_3 = \lambda_4 = 0.5$ and $\lambda_5 = 0.1$ are the weighting coefficients.

During training, discrepancies between the feature maps of the student and teacher networks are unavoidable. In the mid-to-late stages of training, such discrepancies may also affect the learning progress of the student network. We hope that the teacher network guides the student network in the early training stages, while gradually reducing its influence in later stages, which leads the student network to a self-learning phase. Therefore, we add a decay function $D$ to the $KD$ loss, which is defined as 
\begin{equation}
D=1\times{d}^{\lfloor {epoch/f} \rfloor},
\end{equation}
where $d$ is the decay factor, $f$ is the decay frequency, while $epoch$ represents the current training epoch. It decreases the contribution to the total training loss progressively with the epoch increasing. The final total training loss is defined as follows
\begin{equation}
\mathcal{L}_{total}(\Theta) = D \cdot \alpha \mathcal{L}_{KD} + \mathcal{L}_{H},
\end{equation}
where $\alpha$ represent the initial contributions of $KD$ loss.

\section{EXPERIMENTS AND RESULTS}
\label{exp}

\subsection{Ablation Study}
We ablate each component step by step to investigate their performance and the effectiveness of the proposed method. All experiments are performed on the Chikusei dataset, with the super-resolution factor of $4$.

\begin{table}[htbp]
\centering
\caption{Ablation study on balancing model performance and computational cost.}
\label{tab:blocks}
\renewcommand{\arraystretch}{1.2} 
\resizebox{\columnwidth}{!}{
\begin{tabular}{c | c c | c c c c c c}
\hline
Blocks & param & flops & MPSNR$\uparrow$ & MSSIM$\uparrow$ & SAM$\downarrow$ & CC$\uparrow$ & RMSE$\downarrow$ & ERGAS$\downarrow$ \\
\hline
4	& 2.810M & 0.698G & 39.7607 & 0.9364 & 2.4479 & 0.9505 & 0.0122 & 5.2899 \\
8	& 3.109M & 0.754G & 40.0476 & 0.9403 & 2.3543 & 0.9535 & 0.0118 & 5.1221 \\
16	& 3.707M & 0.867G & 40.3106 & 0.9438 & 2.2728 & 0.9561 & 0.0114 & 4.9765 \\
32	& 4.904M & 1.092G & 40.5037 & 0.9462 & 2.2268 & 0.9579 & 0.0112 & 4.8643 \\
48	& 6.100M & 1.317G & 40.5590 & 0.9469 & 2.2124 & 0.9584 & 0.0111 & 4.8360 \\
64	& 7.297M & 1.543G & 40.5793 & 0.9473 & 2.2013 & 0.9585 & 0.0111 & 4.8320 \\
\hline
\end{tabular}
}
\end{table}

\noindent\textbf{The number of LKB modules:} Since this paper focuses on the lightweight study, it is essential to balance model performance and computational cost. To this end, we explore the impact of the number of LKB modules on the overall model performance. As can be seen from Table \ref{tab:blocks}, both performance and computational cost increase with the growth of LKB modules. Therefore, we ultimately select a network with 16 stacked LKB modules for our subsequent experiments. This configuration strikes a good balance between computational cost and performance, meeting the needs of a lightweight network.
\begin{table}[htbp]
\centering
\caption{A study on the effectiveness of the decay factor.}
\label{tab:df}
\renewcommand{\arraystretch}{1.2} 
\resizebox{\columnwidth}{!}{
\begin{tabular}{c | c | c c c c c c}
\hline
$d$ & $\alpha$ & MPSNR$\uparrow$ & MSSIM$\uparrow$ & SAM$\downarrow$ & CC$\uparrow$ & RMSE$\downarrow$ & ERGAS$\downarrow$ \\
\hline
1.00 & \multirow{4}{*}{\centering 0.01} & 40.1720 & 0.9420 & 2.3341 & 0.9548 & 0.0116 & 5.0618 \\
0.75 & & 40.2210 & 0.9425 & 2.3102 & 0.9553 & 0.0116 & 5.0272 \\
0.66 & & 40.2351 & 0.9425 & 2.3110 & 0.9554 & 0.0115 & 5.0165 \\
0.50 & & 40.2329 & 0.9425 & 2.3124 & 0.9554 & 0.0115 & 5.0195 \\
\hline
\end{tabular}
}
\end{table}

\noindent\textbf{The decay $d$ of $KD$ loss:} As mentioned in the methodology section \ref{FA}, it is harmful to allow the teacher network to fully supervise the knowledge distillation process on training the student. Therefore, we introduce a decay function $D$ to the $KD$ loss in order to reduce the influence of the teacher in the later stages of training. To validate the effectiveness of this strategy, we fix the initial proportion of $KD$ loss as $\alpha=0.01$, and adjust the value of the decay $d$, to show the impact of the decay function on KD results. From Table \ref{tab:df}, we find that the improvement of performance is negligible when the decay $d$ is set to 1, which means the teacher fully supervises the student during the KD processing. Specifically, the MPSNR value increases only 0.003 dB compared to training of the student network (40.1691) without KD. Meanwhile, the MPSNR value improves by 0.066 dB when the decay factor is set to 0.66, which validates the effectiveness of introducing a decay into $KD$ loss. It is noteworthy that we do not aim to discuss the optimal setting for the decay factor. Instead, we want to demonstrate the effectiveness of this strategy through this ablation study. Therefore, we simply set it to be 0.66, and the decay frequency $f$ is set to once every 10 training epochs.
\begin{table}[htbp]
\centering
\caption{Ablation study on the initial proportion of distillation loss.}
\label{tab:a}
\renewcommand{\arraystretch}{1.2}
\resizebox{\columnwidth}{!}{
\begin{tabular}{c | c | c c c c c c}
\hline
$d$ & $\alpha$ & MPSNR$\uparrow$ & MSSIM$\uparrow$ & SAM$\downarrow$ & CC$\uparrow$ & RMSE$\downarrow$ & ERGAS$\downarrow$ \\
\hline
\multirow{5}{*}{\centering 0.66} & 1.000 & 40.1817 & 0.9423 & 2.3313 & 0.9549 & 0.0116 & 5.0626 \\
& 0.100 & 40.2071 & 0.9425 & 2.3208 & 0.9552 & 0.0115 & 5.0458 \\
& 0.050 & 40.2294 & 0.9427 & 2.3117 & 0.9554 & 0.0115 & 5.0253 \\
& 0.010 & 40.2351 & 0.9425 & 2.3110 & 0.9554 & 0.0115 & 5.0165 \\
& 0.005 & 40.2231 & 0.9424 & 2.3136 & 0.9553 & 0.0116 & 5.0219 \\
\hline
\end{tabular}
}
\end{table}

\noindent\textbf{The initial contribution $\alpha$ of $KD$ loss:} The initial contribution $\alpha$ of $KD$ loss also influences the performance of KD. Here, we fix the decay $d$ of $KD$ loss to $0.66$ in order to find an appropriate $\alpha$ for $KD$ loss. From Tab.~\ref{tab:a}, we find that the student network achieves the best performance with an MPSNR value of $40.2351$, when $\alpha$ is set to $0.01$. Therefore, we set $\alpha$ to $0.01$ for all subsequent distillation experiments.
\begin{table}[htbp]
\centering
\caption{The ablation study on the composition of distillation loss.}
\label{tab:loss}
\renewcommand{\arraystretch}{1.2} 
\resizebox{\columnwidth}{!}{
\begin{tabular}{c c c | c c c c}
\hline
$\mathcal{L}_{\text{sam}}$ & $\mathcal{L}_{\text{grad}}$ & $\mathcal{L}_{\text{cos}}$ & MPSNR$\uparrow$ & MSSIM$\uparrow$ & SAM$\downarrow$ & ERGAS$\downarrow$ \\
\hline
\checkmark & \checkmark & \checkmark  & 40.2351 & 0.9425 & 2.3110 & 5.0165 \\
\checkmark & \checkmark &  & 40.2003 & 0.9422 & 2.3174 & 5.0378 \\
\hline
\end{tabular}
}
\end{table}

\noindent\textbf{Ablation study on the $KD$ loss:} The proposed $KD$ loss is composed of $SAM$ loss $\mathcal{L}_{sam}$, gradient loss $\mathcal{L}_{grad}$ and  cosine similarity loss $\mathcal{L}_{cos}$. As shown in Table~\ref{tab:loss}, the combination of $\mathcal{L}_{sam}$, $\mathcal{L}_{grad}$, and $\mathcal{L}_{cos}$ achieves the better distillation performance. Specifically, compared to the case without cosine loss, the MPSNR value increased by 0.035 dB. These results validate our hypothesis described in the methodology section: adding a loss term to further constrain the directional consistency of spectral vectors in feature maps is effective.

\subsection{Datasets and Experimental Setup}
In this section, we conduct a detailed analysis and evaluation of our model's performance on three publicly available hyperspectral image datasets including Chikusei, Pavia Center, and Houston 2018 datasets. We compare the proposed model with five state-of-the-art methods in the SHSR domain, such as VDSR \cite{vdsr}, 3DFCNN \cite{3dfcnn}, SSPSR \cite{sspsr}, MSDformer \cite{msdformer}, and ESSAformer \cite{essaformer}. VDSR and 3DFCNN have comparable amounts of parameters and computational costs. SSPSR, MSDformer, and ESSAformer have significantly higher amount of parameters than ours. To ensure fairness, we maintain the settings of these comparison methods as described in their original papers. We conducted experiments with super-resolution scaling factors of $r=4$ and $r=8$ on all datasets. Our models are trained from scratch with PyTorch and the Adam optimizer. Drop path regularization is only applied in training. The initial learning rate is $2 \times 10^{-3}$, which is gradually decreased to a minimum of $2 \times 10^{-4}$. All experiments are conducted on NVIDIA RTX 4090 GPUs.
\begin{figure*}[ht]
	\centering
	\includegraphics[scale=0.26]{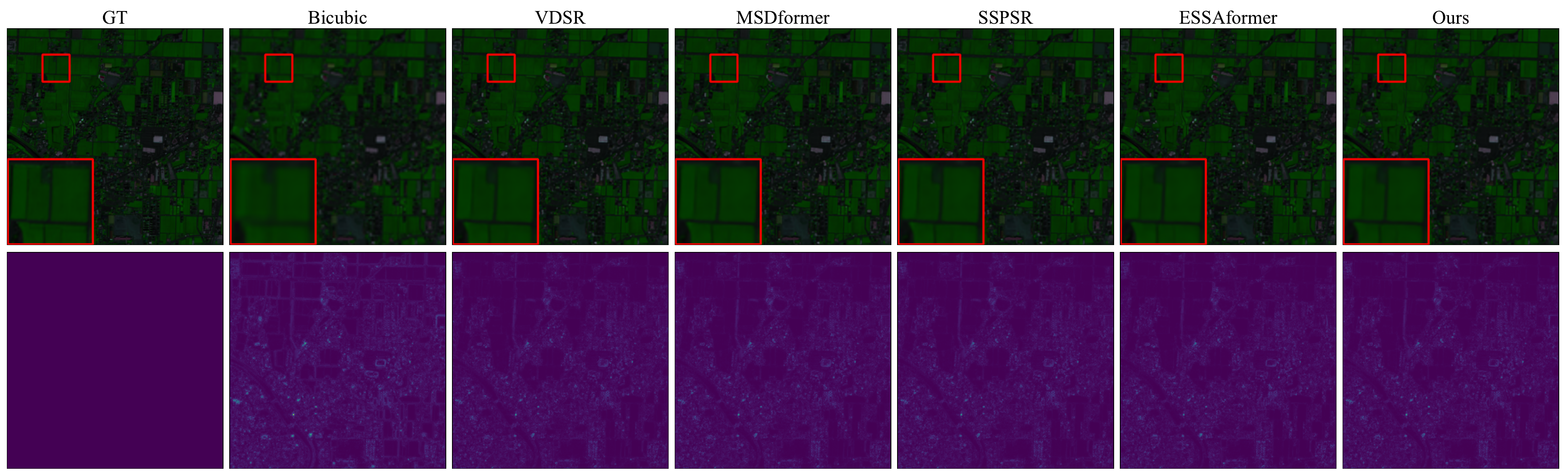}
    \caption{Reconstructed composite images (the first row) and the error maps (the second row) of one test hyperspectral image in Chikusei dataset with spectral bands 31-98-61 as R-G-B with upsampling factor $r = 4$.}
	\label{chix4}
\end{figure*}

\begin{table*}[htbp]
\centering
\caption{AVERAGE QUANTITATIVE COMPARISONS OF NINE DIFFERENT APPROACHES OVER FOUR TESTING IMAGES FROM CHIKUSEI DATASET WITH RESPECT TO SIX PQIS.}
\label{tab:chikusei_results}
\begin{tabular}{c c | c c | c c c c c c}
\hline\rule{0pt}{2.5ex}
Ratio & \centering Method        & Param   & Flops   & MPSNR$\uparrow$  & MSSIM$\uparrow$  & SAM$\downarrow$   & CC$\uparrow$     & RMSE$\downarrow$   & ERGAS$\downarrow$     \\
\hline\hline\rule{0pt}{2.5ex}
\multirow{9}{*}{\centering 4} & \centering Bicubic    & -       & -       & 37.6377 & 0.8953 & 3.4040 & 0.9212 & 0.0155 & 6.7563 \\
                  & \centering VDSR\cite{vdsr}       & \underline{1.917M}  & \underline{0.491G}  & 38.6684 & 0.9227 & 2.6242 & 0.9364 & 0.0135 & 6.1510 \\
                  & \centering 3DFCNN\cite{3dfcnn}     & 3.378M  & 0.864G  & 38.0440 & 0.9115 & 3.8237 & 0.9249 & 0.0144 & 6.6512 \\
                  & \centering SSPSR\cite{sspsr}      & 13.564M & 43.484G & 40.3711 & 0.9444 & 2.3259 & 0.9567 & 0.0114 & 4.9145 \\
                  & \centering MSDformer\cite{msdformer}  & 15.527M & 13.109G & 40.1000 & 0.9408 & 2.3991 & 0.9541 & 0.0118 & 5.0799 \\
                  & \centering ESSAformer\cite{essaformer} & 11.641M & 52.135G & 40.6626 & 0.9480 & 2.2707 & 0.9592 & 0.0110 & 4.7884 \\
[0.5ex]\cline{2-10}\rule{0pt}{2.5ex}                  & \centering LKCA-Net       & 3.707M  & 0.867G  & 40.3106 & 0.9438 & 2.2728 & 0.9561 & 0.0114 & 4.9765 \\
                  & \centering LKCA-LR    & 1.643M  & 0.338G  & 40.1691 & 0.9418 & 2.3219 & 0.9547 & 0.0116 & 5.0552 \\
                  & \centering LKCA-KD    & \textbf{1.643M}  & \textbf{0.338G}  & 40.2351 & 0.9425 & 2.3110 & 0.9554 & 0.0115 & 5.0165 \\
[0.5ex]\hline\rule{0pt}{2.5ex}
\multirow{9}{*}{\centering 8} & \centering Bicubic    & -       & -       & 34.5049 & 0.8069 & 5.0436 & 0.8314 & 0.0224 & 4.8488 \\
                  & \centering VDSR\cite{vdsr}       & \underline{5.456M}  & \underline{1.397G}  & 34.9899 & 0.8288 & 4.3905 & 0.8483 & 0.0209 & 4.6232 \\
                  & \centering 3DFCNN\cite{3dfcnn}     & 7.531M  & 1.926G  & 34.6766 & 0.8114 & 5.3196 & 0.8360 & 0.0216 & 4.7927 \\
                  & \centering SSPSR\cite{sspsr}      & 15.925M & 122.891G  & 35.9117 & 0.8563 & 3.9619 & 0.8792 & 0.0190 & 4.1237 \\
                  & \centering MSDformer\cite{msdformer}  & 17.601M & 28.398G & 35.6897 & 0.8481 & 4.0628 & 0.8723 & 0.0195 & 4.2131 \\
                  & \centering ESSAformer\cite{essaformer} & 14.149M & 205.593G  & 36.0120 & 0.8585 & 3.9826 & 0.8817 & 0.0188 & 4.0767 \\
[0.5ex]\cline{2-10}\rule{0pt}{2.5ex}                  & \centering LKCA-Net       & 10.798M & 2.679G  & 35.8811 & 0.8543 & 3.8639 & 0.8777 & 0.0190 & 4.1323 \\
                  & \centering LKCA-LR    & 2.540M  & 0.565G  & 35.8286 & 0.8521 & 3.9353 & 0.8763 & 0.0192 & 4.1571 \\
                  & \centering LKCA-KD    & \textbf{2.540M}  & \textbf{0.565G}  & 35.8466 & 0.8523 & 3.9341 & 0.8769 & 0.0191 & 4.1468 \\
[0.5ex]\hline\rule{0pt}{2.5ex}
\end{tabular}
\end{table*}

To evaluate the performance of the model, we employ comprehensive evaluation metrics in the SHSR domain, including peak signal-to-noise ratio (PSNR), structural similarity index (SSIM), spectral angle mapper (SAM), correlation coefficient (CC), root mean square error (RMSE), and the relative global dimensional synthesis error (ERGAS). PSNR, SSIM, and RMSE are typically used to evaluate the reconstruction accuracy of natural images, calculated on single-channel images. To well compare super-resolution performance, we record the average values across all spectral bands. Meanwhile, CC, SAM, and ERGAS are commonly used for hyperspectral image fusion tasks.

\subsection{Experimental Results on Chikusei Dataset}
The Chikusei dataset was captured using the Headwall Hyperspec VNIR-C imaging sensor, in an urban area of Chikusei City, Ibaraki Prefecture, Japan. It contains 128 spectral bands, with a spectral range of $363 \, \text{nm}$ to $1018 \, \text{nm}$, a total image size of $2517 \times 2335$ pixels, and a ground sampling distance (GSD) of $2.5 \, \text{m}$.

As suggested in SSPSR \cite{sspsr}, we crop the original image to the central region containing valid information, with a size of $2304 \times 2048 \times 128$. Specifically, we crop four non-overlapping regions of size $512 \times 2048 \times 128$ from the top of the dataset for testing. The remaining regions are randomly sampled for training (10\% of the data was selected for validation). For a super-resolution scaling factor of $r = 4$, we crop patches of size $64 \times 64$, with 32 pixels overlap. For $r = 8$, we crop patches of size $128 \times 128$, with 64 pixels overlap. These small patches are used as high-resolution (HR) hyperspectral images, which serve as the ground truth (GT). The corresponding low-resolution (LR) hyperspectral images are generated by downsampling these HR images by factors of 4 and 8 using Bicubic interpolation to ensure consistent scaling factors.

\begin{figure*}[ht]
	\centering
	\includegraphics[scale=0.26]{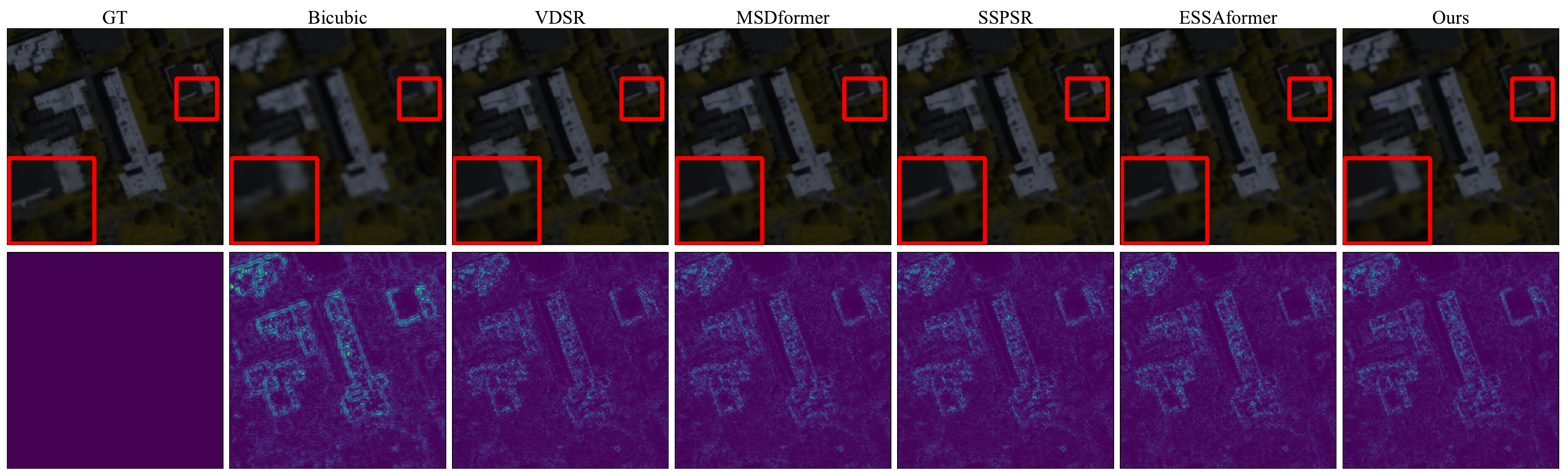}
    \caption{Reconstructed composite images (the first row) and the error maps (the second row) of one test hyperspectral image in Houston 2018 dataset with spectral bands 29-26-19 as R-G-B with upsampling factor $r = 4$.}
	\label{houx4}
\end{figure*}
\begin{table*}[htbp]
\centering
\caption{AVERAGE QUANTITATIVE COMPARISONS OF NINE DIFFERENT APPROACHES OVER EIGHT TESTING IMAGES FROM HOUSTON DATASET WITH RESPECT TO SIX PQIS.}
\label{tab:houston_results}
\begin{tabular}{c c | c c | c c c c c c}
\hline\rule{0pt}{2.5ex}
Ratio & \centering Method        & Param   & Flops   & MPSNR$\uparrow$  & MSSIM$\uparrow$  & SAM$\downarrow$   & CC$\uparrow$     & RMSE$\downarrow$   & ERGAS$\downarrow$     \\
\hline\hline\rule{0pt}{2.5ex}
\multirow{9}{*}{\centering 4} & \centering Bicubic    & -       & -       & 43.0272 & 0.9613 & 2.5453 & 0.9741 & 0.0086 & 2.9085 \\
                  & \centering VDSR\cite{vdsr}       & \underline{1.268M}  & 0.324G  & 45.7846 & 0.9787 & 1.8494 & 0.9858 & 0.0061 & 2.0926 \\
                  & \centering 3DFCNN\cite{3dfcnn}     & \textbf{1.134M}  & \underline{0.290G}  & 44.1961 & 0.9711 & 2.5401 & 0.9788 & 0.0072 & 2.4900 \\
                  & \centering SSPSR\cite{sspsr}      & 13.011M & 20.538G & 46.4067 & 0.9815 & 1.7393 & 0.9877 & 0.0058 & 1.9538 \\
                  & \centering MSDformer\cite{msdformer}  & 13.536M & 7.675G  & 46.5412 & 0.9819 & 1.7188 & 0.9880 & 0.0057 & 1.9273 \\
                  & \centering ESSAformer\cite{essaformer} & 11.272M & 51.333G & 47.0629 & 0.9841 & 1.6890 & 0.9893 & 0.0053 & 1.8070 \\
[0.5ex]\cline{2-10}\rule{0pt}{2.5ex}                  & \centering LKCA-Net       & 2.138M  & 0.466G  & 46.5215 & 0.9819 & 1.7057 & 0.9879 & 0.0057 & 1.9251 \\
                  & \centering LKCA-LR    & 1.364M  & 0.268G  & 46.4335 & 0.9817 & 1.7286 & 0.9877 & 0.0057 & 1.9449 \\
                  & \centering LKCA-KD    & 1.364M  & \textbf{0.268G}  & 46.4663 & 0.9818 & 1.7244 & 0.9878 & 0.0057 & 1.9372 \\
[0.5ex]\hline\rule{0pt}{2.5ex}
\multirow{9}{*}{\centering 8} & \centering Bicubic    & -       & -       & 38.1083 & 0.8987 & 4.6704 & 0.9177 & 0.0152 & 2.5615 \\
                  & \centering VDSR\cite{vdsr}       & \underline{2.461M}  & \underline{0.630G}  & 39.4723 & 0.9188 & 3.7640 & 0.9381 & 0.0128 & 2.1787 \\
                  & \centering 3DFCNN\cite{3dfcnn}     & 2.826M  & 0.723G  & 38.4889 & 0.9069 & 4.8248 & 0.9239 & 0.0140 & 2.3980 \\
                  & \centering SSPSR\cite{sspsr}      & 15.372M & 62.706G & 40.0255 & 0.9285 & 3.5196 & 0.9457 & 0.0120 & 2.0420 \\
                  & \centering MSDformer\cite{msdformer}  & 15.611M & 18.717G & 39.9318 & 0.9261 & 3.5327 & 0.9442 & 0.0121 & 2.0592 \\
                  & \centering ESSAformer\cite{essaformer} & 13.780M & 202.526G  & 40.5941 & 0.9354 & 3.3151 & 0.9519 & 0.0112 & 1.9075 \\
[0.5ex]\cline{2-10}\rule{0pt}{2.5ex}                  & \centering LKCA-Net       & 4.797M  & 1.145G  & 40.3546 & 0.9316 & 3.3421 & 0.9491 & 0.0116 & 1.9657 \\
                  & \centering LKCA-LR    & 1.700M  & 0.353G  & 40.2564 & 0.9303 & 3.3922 & 0.9480 & 0.0117 & 1.9889 \\
                  & \centering LKCA-KD    & \textbf{1.700M}  & \textbf{0.353G}  & 40.2936 & 0.9307 & 3.3783 & 0.9485 & 0.0117 & 1.9807 \\
[0.5ex]\hline\rule{0pt}{2.5ex}
\end{tabular}
\end{table*}

Tab. \ref{tab:chikusei_results} shows the comparison results of LKCA-Net with five other SHSR methods. Additionally, we record the performance of the low-rank network (LKCA-LR) and the network with feature alignment (LKCA-KD), to verify the feasibility of the proposed low-rank approximation and feature alignment methods. We can see that LKCA-Net, LKCA-LR, and LKCA-KD involve significantly fewer parameters and computational costs compared to heavy models like MSDformer, yet achieve comparable or even better performance. For example, in the case of $r=4$, LKCA-Net achieves a MPSNR of 40.31dB, which is higher than the 40.10dB of MSDformer, with only 3.71M parameters and 0.867G FLOPs (4 times and 15 times lower compared to MSDformer). This demonstrates that the proposed LKCA-Net can achieve a good balance between performance and complexity. The low-rank version LKCA-LR still surpasses MSDformer at 40.17dB MPSNR with 9 times fewer parameters and 39 times fewer FLOPs. The performance of LKCA-LR is slightly lower than LKCA-Net, but still better than MSDformer, demonstrating the effectiveness of the proposed low-rank approximation method. The feature alignment version LKCA-KD achieves a similar performance to LKCA-Net, demonstrating the effectiveness of the proposed feature alignment method. We have similar conclusions for $r=8$.

Figure \ref{chix4} shows the visual results on one test hyperspectral image in the Chikusei dataset, including the ground truth, bicubic interpolation, the lightweight networks including VDSR, the large network MSDformer, SSPSR and ESSAformer, and our method LKCA-KD. From the visual results, it can be observed that LKCA-KD significantly outperforms VDSR. Moreover, LKCA-KD demonstrates a recovery capability for fine-grained textures and coarse-grained structures  (as highlighted in the red-boxed areas) that is very close to the performance of the best-performing ESSAformer, which clearly demonstrates the competitiveness of our method.

\subsection{Experimental Results on Houston Dataset}

\begin{figure*}[ht]
	\centering
	\includegraphics[scale=0.26]{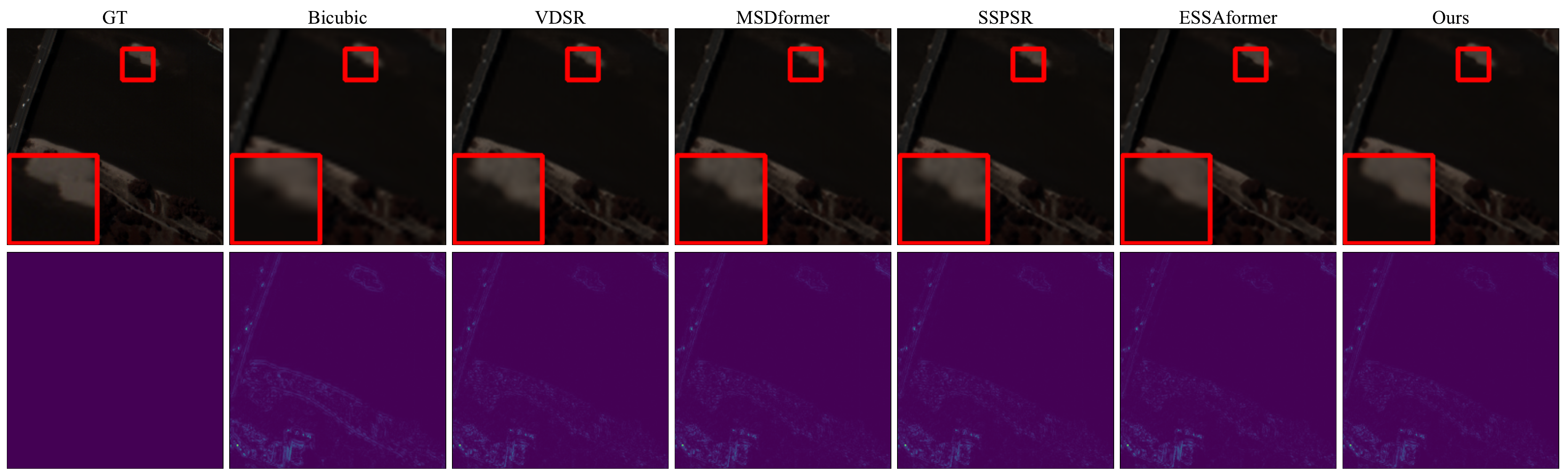}
    \caption{Reconstructed composite images (the first row) and the error maps (the second row) of one test hyperspectral image in Pavia Center dataset with spectral bands 30-20-12 as R-G-B with upsampling factor $r = 4$.}
	\label{pax4}
\end{figure*}
\begin{table*}[htbp]
\centering
\caption{AVERAGE QUANTITATIVE COMPARISONS OF NINE DIFFERENT APPROACHES OVER THREE TESTING IMAGES FROM PAVIA DATASET WITH RESPECT TO SIX PQIS.}
\label{tab:pavia_results}
\begin{tabular}{c c | c c | c c c c c c}
\hline\rule{0pt}{2.5ex}
Ratio & \centering Method        & Param   & Flops   & MPSNR$\uparrow$  & MSSIM$\uparrow$  & SAM$\downarrow$   & CC$\uparrow$     & RMSE$\downarrow$   & ERGAS$\downarrow$     \\
\hline\hline\rule{0pt}{2.5ex}
\multirow{9}{*}{\centering 4} & \centering Bicubic    & -       & -       & 30.6810 & 0.8323 & 6.8835 & 0.9205 & 0.0325 & 7.7000 \\
                  & \centering VDSR\cite{vdsr}        & \underline{1.662M}  & \underline{0.426G}  & 31.5422 & 0.8646 & 6.7023 & 0.9344 & 0.0298 & 6.9838 \\
                  & \centering 3DFCNN\cite{3dfcnn}      & 2.692M  & 0.689G  & 31.3055 & 0.8587 & 7.6825 & 0.9314 & 0.0299 & 7.1894 \\
                  & \centering SSPSR\cite{sspsr}       & 13.385M & 36.394G & 32.2823 & 0.8884 & 6.4033 & 0.9447 & 0.0272 & 6.4407 \\
                  & \centering MSDformer\cite{msdformer}   & 14.908M & 11.413G & 32.3802 & 0.8911 & 6.3422 & 0.9464 & 0.0267 & 6.3706 \\
                  & \centering ESSAformer\cite{essaformer}  & 11.521M & 51.874G & 32.7185 & 0.9014 & 6.2703 & 0.9503 & 0.0255 & 6.1344 \\
[0.5ex]\cline{2-10}\rule{0pt}{2.5ex}                  & \centering LKCA-Net       & 3.197M  & 0.737G  & 32.5504 & 0.8970 & 6.1378 & 0.9488 & 0.0261 & 6.2566 \\
                  & \centering LKCA-LR    & 1.552M  & 0.315G  & 32.5682 & 0.8960 & 6.1520 & 0.9487 & 0.0261 & 6.2432 \\
                  & \centering LKCA-KD    & \textbf{1.552M}  & \textbf{0.315G}  & 32.6104 & 0.8971 & 6.1407 & 0.9491 & 0.0260 & 6.2127 \\
[0.5ex]\hline\rule{0pt}{2.5ex}
\multirow{9}{*}{\centering 8} & \centering Bicubic    & -       & -       & 27.4155 & 0.6782 & 8.3665 & 0.8291 & 0.0472 & 5.5958 \\
                  & \centering VDSR\cite{vdsr}        & \underline{4.482M}  & \underline{1.148G}  & 27.4992 & 0.6796 & 8.4967 & 0.8317 & 0.0467 & 5.5418 \\
                  & \centering 3DFCNN\cite{3dfcnn}      & 6.002M  & 1.535G  & 27.4940 & 0.6715 & 11.9678 & 0.8347 & 0.0461 & 5.5711 \\
                  & \centering SSPSR\cite{sspsr}       & 15.745M & 104.255G  & 28.1251 & 0.7218 & 8.1682 & 0.8526 & 0.0439 & 5.1488 \\
                  & \centering MSDformer\cite{msdformer}   & 16.983M & 25.321G & 27.9797 & 0.7131 & 8.2815 & 0.8494 & 0.0444 & 5.2396 \\
                  & \centering ESSAformer\cite{essaformer}  & 14.029M & 204.596G  & 28.0796 & 0.7253 & 8.2574 & 0.8506 & 0.0440 & 5.1795 \\
[0.5ex]\cline{2-10}\rule{0pt}{2.5ex}                  & \centering LKCA-Net       & 8.847M  & 2.18G   & 28.2250 & 0.7279 & 7.7476 & 0.8551 & 0.0434 & 5.0913 \\
                  & \centering LKCA-LR    & 2.267M  & 0.496G  & 28.2605 & 0.7290 & 7.6483 & 0.8564 & 0.0432 & 5.0701 \\
                  & \centering LKCA-KD    & \textbf{2.267M}  & \textbf{0.496G}  & 28.2340 & 0.7297 & 7.6163 & 0.8553 & 0.0434 & 5.0889 \\
[0.5ex]\hline\rule{0pt}{2.5ex}
\end{tabular}
\end{table*}

The Houston 2018 dataset is part of the IEEE GRSS Data Fusion Contest in 2018, captured by the ITRES CASI 1500 hyperspectral sensor. It covers the campus of the University of Houston and surrounding urban areas in Houston, Texas, USA. The dataset includes $4172 \times 1202$ pixels, spanning 48 spectral bands with a wavelength range of $380 \, \text{nm}$ to $1050 \, \text{nm}$ and a ground sampling distance (GSD) of $1 \, \text{m}$.

We crop eight non-overlapping regions of size $256 \times 256 \times 48$ from the top $512 \times 1024 \times 48$ region of the dataset as high-resolution hyperspectral images for testing. The remaining $512 \times 178 \times 48$ region from the top is discarded due to its insufficient size for testing purposes. The rest of the image ($4172 \times 1202 \times 48$) is randomly cropped into small patches for training (10\% of the training data are reserved for validation). Low-resolution (LR) hyperspectral images are generated by downsampling these small patches using Bicubic interpolation.

Table \ref{tab:houston_results} presents the experimental results on the Houston dataset. Note that Houston is the simplest dataset among the three datasets. The performance of the LKCA-Net significantly surpasses other methods in the same parameter range and demonstrates excellent competitiveness against non-lightweight methods. For $r=4$, LKCA-Net outperforms SSPSR and achieves comparable performance to that of MSDformer. For $r=8$, the performance of LKCA-Net significantly exceeds both SSPSR and MSDformer. 

For both $r=4$ and $r=8$, LKCA-Net exhibits a significantly lower number of parameters and computational cost compared to SSPSR and MSDformer. Specifically, for $r=4$, LKCA-Net's computational cost is less than one percent of that of ESSAformer, and for $r=8$, it is approximately 0.5 percent. Despite the substantial reduction in computational cost, LKCA-Net maintains performance comparable to that of ESSAformer. These findings clearly demonstrate the effectiveness of the performance optimizations introduced by LKCA-Net.

Figure \ref{houx4} shows the visual results on a hyperspectral image from the Houston 2018 test dataset. Similarly, on this dataset, LKCA-KD stands out among lightweight methods, with its ability to recover details comparable to that of ESSAformer. The results demonstrate the effectiveness of our method.

\subsection{Experimental Results on Pavia Dataset}
The Pavia Centre dataset was captured using the Reflective Optics System Imaging Spectrometer (ROSIS) during a flight campaign conducted in 2001 over the central region of Pavia, northern Italy. The dataset consists of $1096 \times 1096$ pixels and 102 spectral bands with a wavelength range from $430 \, \text{nm}$ to $860 \, \text{nm}$ and a ground sampling distance of $1.3 \, \text{m}$. In the Pavia Centre dataset, regions without meaningful information in the hyperspectral image are removed, leaving a final effective region of size $1096 \times 715 \times 102$. Specifically, we crop the top $224 \times 715 \times 102$ region of the dataset into three $224 \times 224 \times 102$ non-overlapping hyperspectral patches for testing. The rightmost area, which does not meet the required size for testing patches, is discarded, while the remaining regions are cropped into small blocks for training (with $10\%$ of the training data used as validation samples).

Table \ref{tab:pavia_results} presents the experimental results on the Pavia dataset. With $r = 4$, the performance of LKCA-KD surpasses all methods except the largest model, ESSAformer. Actually, the performance of LKCA-KD is very close to that of ESSAformer, only 0.11 dB lower in MPSNR but involves 13\% of the parameters and 0.6\% of the computational cost. The same conclusion can be drawn for $r = 8$. The results on the Pavia dataset further demonstrate the effectiveness of the proposed LKCA-Net and the feature alignment strategy. Moreover, we can see that the LKCA-KD outperforms the original LKCA-Net slightly. The reason is that the Pavia dataset is the most challenging dataset among the three datasets due to the smallest number of training samples, as a result, the fewer parameters of LKCA-KD have lower overfitting risk. 

Figure \ref{pax4} shows the visual results on a hyperspectral image from the Pavia Center test dataset. As indicated by the red-boxed region, it can be observed that our method exhibits higher smoothness in restoring the edge details of the image, even outperforming the heavyweight networks MSDformer and SSPSR. This highlights the competitive advantage of our approach when handling complex datasets.

\section{CONCLUSIONS}
\label{con}

This paper proposed a novel lightweight single hyperspectral image super-resolution method, which included a lightweight backbone (LKCA-Net) and a low-rank approximation upsampling layer. We designed backbone of CNN architecture, which ensured multi-scale capabilities, large receptive fields, and channel calibration while maintaining parameter and computational efficiency. It effectively enhanced feature extraction performance and provided a new benchmark for lightweight SHSR tasks. For the upsampling layer, we were the first to argue the upsampling layer is a key bottleneck in lightweight SHSR. By confirming its low-rank characteristics, we proposed a sampling low-rank approximation and a feature alignment strategy to optimize its performance. Experimental results demonstrated that our method achieved excellent performance on three public hyperspectral datasets. Compared to state-of-the-art methods, our approach achieved competitive performance and speedups of several dozen to even hundreds of times.

\section{Acknowledgment}
We would like to express our sincere appreciation to the anonymous reviewers.

\bibliographystyle{IEEEtran}
\bibliography{ref}

\vfill

\end{document}